\documentclass[10pt,journal,compsoc]{IEEEtran}

\usepackage{amsmath}
\usepackage{algorithm}
\usepackage{graphicx}
\usepackage{tikz}
\usepackage{alltt}
\usepackage{epstopdf}
\usepackage[bookmarksnumbered,unicode]{hyperref}

\usepackage{algpseudocode}

\newcommand{\tuple}[1]{\(\langle\)#1\(\rangle\)}
\newcommand{\tupleM}[1]{\langle#1\rangle}

\newcounter{algkmeans}
\makeatletter
\newenvironment{algkmeans}[1][htb]{%
  \let\c@algorithm\c@algkmeans
  \renewcommand{\ALG@name}{Algorithm}
  
  \begin{algorithm}[#1]%
  }{\end{algorithm}
}
\makeatother

\newcounter{algpagerank}
\makeatletter
\newenvironment{algpagerank}[1][htb]{%
  \let\c@algorithm\c@algpagerank
  \renewcommand{\ALG@name}{Algorithm}
  
  \begin{algorithm}[#1]%
  }{\end{algorithm}
}
\makeatother

\ifCLASSOPTIONcompsoc
  \usepackage[nocompress]{cite}
\else
  \usepackage{cite}
\fi

\begin{document}
%
\title{A New Framework for Expressing, Parallelizing and Optimizing Big
Data Applications}
%
%
%
%

\author{Anne~Hommelberg,
        Bart~van~Strien,
        Kristian~F.~D.~Rietveld,
        and~Harry~A.~G.~Wijshoff
\IEEEcompsocitemizethanks{
\IEEEcompsocthanksitem Extension of conference papers published at
the International Conference on Parallel Processing
Workshops (ICPPW)~\cite{vstrien-2017} and the ACM Conference on
Computing Frontiers~\cite{hommelberg-cf19}.%
\IEEEcompsocthanksitem A. Hommelberg, B. van
Strien, K. F. D. Rietveld and H. A. G. Wijshoff are with LIACS, Leiden
University, Leiden, The Netherlands.}
}

\markboth{Hommelberg \MakeLowercase{\textit{et al.}}: A New Framework for
Expressing, Parallelizing and Optimizing Big Data Applications}{}
%

\IEEEtitleabstractindextext{%
\begin{abstract}
The Forelem framework was first introduced as a means to optimize database
queries using optimization techniques developed for compilers. Since its
introduction, Forelem has proven to be more versatile and to be applicable
beyond database applications. In this paper we show that the original
Forelem framework can be used to express and optimize Big Data applications,
more specifically: k-Means clustering and PageRank, resulting in
automatically generated implementations of these applications. These
implementations are more efficient than state-of-the-art, hand-written MPI
C/C++ implementations of k-Means and PageRank, as well as significantly
outperform state-of-the-art Hadoop implementations.
\end{abstract}

}

\maketitle

\IEEEdisplaynontitleabstractindextext
\IEEEpeerreviewmaketitle
\IEEEraisesectionheading{\section{Introduction}\label{sec:introduction}}

\IEEEPARstart{W}{hen}
the Forelem framework was first introduced, it addressed the optimization of
(embedded) database queries together with the wrapping C/C++ code and its
associated API layer. In order to do this, database queries were
automatically transformed into explicit loop structures (forelem loops),
which together with the wrapping C/C++ code allowed for integral
optimization. This integral optimization led to significant performance
improvements of database applications \cite{tUPLintro}.

In order to allow for this integral optimization in Forelem, database
tables were translated into reservoirs of tuples and database queries were
translated into forelem loop structures, which iterate over a reservoir of
tuples. To ensure a versatile and flexible optimization
process, each iteration of the forelem loop body was assumed to be
tuple-based and atomic, allowing the specification to be inherently
parallel. For database applications this restriction proved to be very
natural, but surprisingly this restriction has also proven to be very
natural for other applications.  In essence this restriction yields a
program specification which is free from common artifacts, like explicit
data and loop structures, and associated data dependencies.  As a result,
this tuple-based and inherently parallel Forelem specification forces the
program design for many applications to be reconsidered, reimplemented and
broken down to their essence. This results in a more flexible, versatile and
parallel implementation of the original algorithm.  On top of this, the
Forelem framework generates data structures automatically at the end of the
compile chain, so that specific characteristics of the applications can be
taken into account.

Previous work has already shown that the Forelem framework was successful in
optimizing (sparse) matrix computations \cite{tUPL2} and LU factorization
\cite{tUPL-LUfactorization}. In this paper, we investigate the usefulness of
the Forelem framework for Big Data Applications, more specifically, k-Means
and PageRank. For both algorithms, we first describe
how the Forelem specification is derived. Second, we demonstrate how through
the application of simple transformations, different implementations can be
automatically obtained.

The derived implementations are benchmarked with both Hadoop and C/C++ MPI
implementations. For k-Means clustering, the Hadoop implementation is used
that was developed for the scalable machine learning and datamining project
Apache Mahout \cite{mahout}. As a C/C++ MPI implementation a code originally
developed by W.-K.  Liao from Northwestern University~\cite{mpiNW} is
selected. For PageRank, an implementation using Hadoop MapReduce from the
Pegasus~\cite{Kang09} project is used. The evaluated C/C++ MPI
implementation is taken from Indiana University's Parallel
BGL~\cite{BGL}. We show that using the Forelem
framework implementations of k-Means clustering and PageRank are
automatically derived, that improve the C/C++ MPI implementation can
outperform the Hadoop implementations.

This paper is organized as follows. Related work is discussed in
\autoref{sec:related-work}. In \autoref{sec:forelem} the Forelem framework
is described. \autoref{sec:derivation} discusses how Forelem specifications
are derived for the k-Means clustering and PageRank algorithms.
\autoref{sec:transformations} describes a number of transformations and
explains how these are applied to the initial Forelem specifications of the
k-Means and PageRank to derive different implementations. The performance of
the different implementations is evaluated in \autoref{sec:evaluation}.
\autoref{sec:conclusions} concludes the paper.

\section{Related work}
\label{sec:related-work}
As the Forelem framework is based on atomic, tuple-based operations, it may
appear similar to Linda, the tuple space coordination model \cite{linda}.
Linda was introduced by David Gelernter in the eighties using tuples as a
basic operation for coordination and communication of parallel processes.
The basic enabler for these operations was the fact that all tuples were
stored in a physical, shared, virtual, associative memory. All these
operations retrieve tuples from this memory, operate on these tuples, and
store the result tuple back in this memory. Unlike this approach, the
Forelem framework does not make any assumption on where these tuples are
stored, instead tuples are a conceptual notion and act as a placeholder for
the software optimization process of Forelem. It is this difference which
enables Forelem to automatically generate data structures and their mapping
into physical memory. Whereby Linda suffered from performance limitations
thereby hindering the widespread use for high performance parallel
applications, the Forelem framework is far more versatile allowing highly
efficient implementations to be derived for many applications.

As the Forelem framework is tuple-based, there is a natural connection to
the concept of Dataflow computing. Dataflow has been a major topic in
computer architecture research in the seventies and early eighties
\cite{dataflow}. Dataflow computing is token based and at runtime these
tokens are matched and computed on. Several Dataflow computing architectures
were proposed, both for the storage of tokens as well as the
matching unit. For the token storage mostly a content addressable memory was
foreseen. As with Linda, Dataflow computing suffered from performance issues
for general use and its application, although influential, was limited to
specific areas in computer hardware and software design. As such the Forelem
framework can be seen as a generalization of Dataflow computing, enabling a
full optimization chain for general applications.

\section{Forelem}
\label{sec:forelem}
Within Forelem operations are tuple-based and atomic, and are organized by
the use of two different loop structures: the forelem and whilelem loops.
Both structures iterate over the tuples in a tuple reservoir. These tuple
reservoirs are neither physical nor virtual, but are defined on a conceptual
level without specifying any order in which tuples are stored. Tuples
$\texttt{t}$ contain either data fields or index (integer) fields:
\small
\begin{alltt}
\centering \tuple{DATA, I, J}
\end{alltt}
\normalsize
Index fields can be used for reference to other tuples (think of indices in
database tables) or can be used to address data stored in shared spaces. As
with the tuple reservoir, these shared spaces consist only at a conceptual
level. With each shared space $\texttt{A}$ an affine address function
$\texttt{F}_\texttt{A}$ is associated which maps tuple indices to a unique
``location'' in this shared space, in which data can be stored. For the
purpose of readability, in this paper we use simple array notation
$\texttt{[...]}$ to denote these address functions. So, for each tuple
$\texttt{t}:$ $\texttt{A}\texttt{[F}_{\texttt{A}}\texttt{[t]]}$ denotes a
unique address in the shared space $\texttt{A}$. Note that actual data
structures are automatically generated and optimized by the Forelem
transformation engine without involvement of the programmer, therefore this
notation of array indexing should not be confused with actual array data
structures.

The forelem loop from which the framework derives its name, traverses all
tuples in a given tuple reservoir, performing the calculation specified in
the loop body. It executes the loop body, in an arbitrary order, and exactly
once for each tuple. For example, the following forelem loop computes the
product of two possibly sparse matrices located in shared spaces
$\texttt{A}$ and $\texttt{B}$: 
\small
\begin{alltt}
\textbf{forelem} (\tuple{i,j,k} \(\in\) X) 
   C[i,j] += A[i,k] * B[k,j]
\end{alltt}
\normalsize
Here the tuple reservoir $\texttt{X}$ contains tuples for every $\texttt{i}$, $\texttt{j}$, and $\texttt{k}$ for which $\texttt{A}_\texttt{i,k} \neq 0$ and $\texttt{B}_\texttt{k,j} \neq 0$. 
Note that although this closely resembles a C-like implementation of dense matrix $\texttt{A*B}$ multiplication, this specification is different although array notation is used to indicate the accesses to shared spaces $\texttt{A}$, $\texttt{B}$ and $\texttt{C}$, see above. 

The whilelem loop is an extension of the forelem loop, which continues to execute the loop body for different tuples until all of the tuples in the reservoir result in no-op operations.  The loop body may consist of one or more if-statements, guarding the execution of the tuple operations. An example of a whilelem loop is shown below, which sorts the elements in a given shared space $\texttt{A}$:
\small
\begin{alltt}
\textbf{whilelem} (\tuple{i,j} \(\in\) T) 
   \textbf{if} (A[i] > A[j])
      swap(A[i],A[j])
\end{alltt}
\normalsize

\noindent
A possible choice for the tuple reservoir $\texttt{T}$ would contain all
tuples $\langle \texttt{i,j}\rangle$ for which $0 \leq \texttt{i} <
\texttt{j} < \texttt{N}$ with $\texttt{N}$ the total number of elements to
be sorted. However, note that a smaller reservoir which contains only
$\langle \texttt{i,j}\rangle$ such that $\texttt{i} = \texttt{j}-1$ also
suffices, in which case the resulting implementation closely resembles
bubblesort. In fact, by choosing a specific reservoir and order in which
tuples are scheduled, many existing sorting algorithms can easily be derived
from this specification.

Note that neither of the loops specifies a specific order in which the
tuples are traversed. All tuple operations are assumed to be atomic, i.e.,
without interference by the execution of other tuples. Also both loop
structures are inherently parallel, so tuples can be visited in any order
and in parallel. For the whilelem loop structure a tuple can even be
executed multiple times in a row. The actual scheduling of tuple selection
relies on Just Scheduling \cite{just} (not to be confused with Just-in-Time
Scheduling) on which we will not elaborate further in this paper as it is
not directly relevant to the content of this paper and because of page
limitations.  The reader is referred to a forthcoming paper which describes
the formal aspects of this type of scheduling. 

Once a specification is given, several transformations are applied to derive
different implementations \cite{tUPLCPC}. During the code generation
process, the data structures used for the shared spaces will be derived
automatically. Examples of these transformations will be given in
\autoref{sec:transformations}, when deriving the implementations of the
k-Means clustering algorithm and the PageRank algorithm.

\section{Deriving a Forelem Specification}
\label{sec:derivation}
A Forelem specification captures the essence of an algorithm, yielding
a specification free from common artifacts like explicit data and loop
structure and associated data dependencies. This section describes how such
specifications are derived for the k-Means clustering and PageRank
algorithms. The derived specifications are used in the next section to
demonstrate how different implementations can be generated through the
application of basic transformations.

\subsection{k-Means Clustering}
\label{sec:kmeans}
The k-Means clustering algorithm divides a given set of data points of dimension $d$ into $k$ clusters. The number $k$ is specified beforehand by the user. To start, the algorithm first initializes the $k$ cluster centers. This can be done in various ways. A standard distribution consists of randomly assigning data points to one of the $k$ clusters, then calculating the mean of the assigned data points to obtain the cluster center. 

The algorithm consists of several iterations. During each iteration the algorithm loops over each data point, calculating the Euclidean distance to each cluster center. After each iteration, the data points are assigned to the cluster whose center was closest. After reassigning all data points, the cluster centers are set to the mean of all data points that were assigned to this cluster during this iteration:
 
\small
\begin{alltt}
change = \textbf{true}
\textbf{while} (change) \{
  change = \textbf{false} //assume there is no change
  //reassign data points to clusters
  \textbf{for} (x = 1 to N) \{
    mindist = LARGE
    \textbf{for} (m = 1 to k) 
      \textbf{if} (dist(COORDS[x],M_COORDS[m])
          <= mindist) \{
        a = m
        mindist = dist(COORDS[x],M_COORDS[m])
      \}
    \textbf{if} (a != M[x]) \{
      M[x] = a
      change = true 
    \} 
  \}  
  //recalculate cluster centers
  \textbf{if} (change) \{
    \textbf{for} (m = 1 to k) \{
      mean, count = 0
      \textbf{for} (x = 1 to N)
        \textbf{if} (M[x] == m) \{
          mean = mean + COORDS[x]
          count = count + 1
        \}
      M_COORDS[m] = mean/count
    \}
  \}
\}
\end{alltt}
\normalsize
Here \verb+x+ is a data point with \verb+N+ the total number of data points,
\verb+m+ is a cluster with \verb+k+ the total number of clusters,
\verb+dist+ a function that calculates the Euclidean distance, \verb+M[x]+
the cluster data point \verb+x+ is currently assigned to, \verb+COORDS[x]+
the coordinates of data point \verb+x+ and \verb+M_COORDS[m]+ the coordinates of
the cluster center of cluster \verb+m+. Note that \verb+COORDS[x]+ and
\verb+M_COORDS[m]+ are n-dimensional, and any operations involving them,
including the distance function, are in fact n-dimensional operations. 

\subsubsection{Forelem Specification of k-Means}
\label{sec:kmeansforelem}

Recall that the Forelem framework allows ``random'' execution of the (tuple)
operations in any order and for an arbitrary amount of times, in contrast to
the classic implementation of k-Means which explicitly determines the order
of operations. The key difference between the Forelem implementation and
other parallel implementations for k-Means clustering will be that the
computation must be reduced to its core.
To do so we first note that the classic algorithm consists of a main while loop which continues operations until no change is made. This naturally corresponds to using a whilelem loop in the Forelem specification, which by definition terminates as soon as all tuples result in a no-op operation. In fact, the resulting Forelem specification will use only a single outer whilelem loop. 

Next we note that the loop body of the classic algorithm is split into two
separate steps: reassigning the data points and recalculating the cluster
centers. For both steps the classic algorithm contains a 2-dimensional
for-loop, looping over each possible combination of a data point and a cluster.
In the first step, the distance between the data point and the cluster is
compared to the best recorded distance and if this distance is smaller, the
best recorded distance is updated. In the second step, the data point is
then taken into account when recalculating the cluster center, only if the
distance to the given cluster is the best recorded distance. 

For the specification of the whilelem loop these two steps are merged into one, thereby removing the necessary bookkeeping such as the change variable seen in the classic algorithm. After all, these variables do not contribute to the essence of the computation in the classic algorithm. At the same time this merger results in the removal of the artificial barrier between the two steps, resulting in a single whilelem loop in which all steps of the two separate inner for loops are combined into single point operations. In order to ensure that these point operations can be executed in a random fashion and independently of each other, the following observations are used: 
\begin{enumerate}
\item The first step is to capture the whilelem loop body as a simple if-statement: if the distance between a data point $\texttt{x}$ and a cluster $\texttt{m}$ is smaller than the best recorded distance, i.e., the distance to the cluster a data point is currently assigned to (\verb+M[x]+), then we must reassign this data point. If not, then no operation is needed, as is also captured by the use of a boolean recording whether a change occurred in the classic implementation. 
\item The first observation combined with the observation that -- if
$\texttt{M[x] == m}$ then clearly the distance will not be strictly smaller
-- gives us the condition of the if-statement in the whilelem loop: 
\small
\begin{alltt}
\textbf{if} (M[x] != m &&
    dist(COORDS[x],M_COORDS[m])
    < dist(COORDS[x],M_COORDS[M[x]]))
  \{ ... \}
\end{alltt}
\normalsize
where $\texttt{dist}$ calculates the Euclidean distance.
\item As a result, the reassigning of the data point $\texttt{x}$ can then be captured in the body of the if-statement by simply stating $\texttt{M[x] = m}$. 
\end{enumerate}
Therefore, the whilelem loop in the Forelem specification will need to loop
over each combination of a cluster $\texttt{m}$ and a data point
$\texttt{x}$, and take the same steps in the loop body as the classic
algorithm. As a consequence, our reservoir $\texttt{T}$ will contain tuples
$\langle \texttt{m,x}\rangle$. The cluster centers are updated accordingly.

In essence, k-Means clustering can therefore be captured by the following
Forelem whilelem loop given in \autoref{alg:kmeans}.
Here $\texttt{x}$ is a data point and $\texttt{M[x]}$, respectively
$\texttt{m}$, is the cluster $\texttt{x}$ is currently assigned to,
respectively to be assigned to,
$\texttt{dist}$ calculates the Euclidean distance, $\texttt{COORDS[x]}$ is the
coordinates of data point $\texttt{x}$ and $\texttt{M\_COORDS[m]}$ and
$\texttt{M\_SIZE[m]}$ are the cluster center and size of a cluster $\texttt{m}$
respectively. Note again $\texttt{COORDS[x]}$ and $\texttt{M\_COORDS[m]}$ are
n-dimensional and all operations that involve them are in fact n-dimensional
operations, including the distance function $\texttt{dist}$. These
operations have been abbreviated to improve readability.

\begin{algkmeans}
\caption{The initial Forelem specification of k-Means clustering.}
\label{alg:kmeans}
\small
\begin{alltt}
\textbf{whilelem} (\tuple{m,x} \(\in\) T)
  \textbf{if} (M[x] != m && dist(COORDS[x],M_COORDS[m])
      < dist(COORDS[x],M_COORDS[M[x]])) \{
    M_COORDS[M[x]] =
     (M_COORDS[M[x]]*M_SIZE[M[x]] - COORDS[x])
      / (M_SIZE[M[x]] - 1)
    M_SIZE[M[x]] -= 1
    M_COORDS[m] =
     (M_COORDS[m]*M_SIZE[m] + COORDS[x])
      / (M_SIZE[m] + 1)
    M_SIZE[m] += 1
    M[x] = m
  \}
\end{alltt}
\normalsize
\end{algkmeans}

As for the classic algorithm for k-Means, the algorithm converges, albeit
not necessarily to a global optimum.  A proof of the correctness of this
specification and, therefore, its convergence property and its termination,
is given in Appendix A.1.

\subsection{PageRank}
PageRank~\cite{Page99} is an algorithm to rank a set of web pages based on an objective notion of importance.
Intuitively the ranking models random surfers, who, after arriving on a web
page, follow a random link until they stop following successive links.
Once they stop, they pick a random website and continue from there.
Equation~\ref{eqn:pagerank} shows the definition of the PageRank of a
vertex, where the likelihood of stopping is modeled using a constant $0 < d < 1$, also referred to as the damping factor.
\begin{equation}\label{eqn:pagerank}
	PR(v) = \frac{1-d}{|V|} + d\sum_{u \in \text{nbh}^{-}(v)} \frac{PR(u)}{\deg^{+}(u)}
\end{equation}

\noindent
Usually PageRank is calculated in an iterative fashion, where in every
iteration a vertex ``donates'' all its PageRank to its successors, and receives it from its predecessors.
As an example, a vertex with a PageRank of $0.2$, and an outdegree of $5$, yields a rank of $\frac{0.2}{5} = 0.04$ to all its successors.
Additionally, the damping factor is applied to all incoming rank, and the (graph-) constant value $\frac{1-d}{|V|}$ is added.

In~\cite{Brin98} a damping factor of approximately $0.85$ is suggested to
yield the best results. In the rest of this paper we will assume $d=0.85$,
unless stated otherwise.

The de facto standard iterative algorithm for PageRank works by taking the current guess for the PageRank values ($\frac{1}{|V|}$) and calculating the next set of values based on those.
Whenever an iteration results in the same values as the previous generation,
the algorithm has converged to the right values.
The pseudocode for such an implementation is shown here:

\small
\begin{alltt}
\textbf{for} (v = 1 to N)
  PR[v] = 1/N
\textbf{while} (PR != PRold) \{
  PRold = PR
  \textbf{for} (v = 1 to N) \{
    PR[v] = (1-d)/N
    \textbf{for} (u = 1 to N)
      \textbf{if} (edge (u, v) exists)
        PR[v] += d*PRold[u]/Dout[u]
  \}
\}
\end{alltt}
\normalsize

\noindent
Here \texttt{PR} represents the PageRank values of the current generation, \texttt{PRold} the values of the previous generation, \texttt{N} the number of vertices, and \texttt{Dout} the outdegrees of the vertices.

\subsubsection{Forelem Specification of PageRank}
As is the case for k-Means clustering, the PageRank algorithm explicitly
determines an order of operation.  When designing a Forelem specification of
PageRank it is important to reduce the computation such that the basic
operations can be executed independently.

The core computation in PageRank is the value of the sum, shown in Equation~\ref{eqn:pagerank}, and the imposed order is simply to prevent summing the wrong values.
The iterations are intended to ``forget'' the previous calculation, so old
(wrong) values can be removed. In order to reduce the algorithm to its
essence and therefore can be executed independently, the Forelem
implementation keeps track of differences of these updates for each
tuple. Then these differences are used to adjust the PageRank values instead
of each time recalculating the summation.
Implementing this change leads us to the following Forelem specification:
\small
\begin{alltt}
\textbf{whilelem} (w \(\in\) V)
  \textbf{if} (PR[w] != OLD[w]) \{
    \textbf{forelem} (\tuple{u,v} \(\in\) E.u[w])
      PR[v] += d*(PR[w]-OLD[w])/Dout[w]
    OLD[w] = PR[w]
  \}
\end{alltt}
\normalsize
\noindent
Here \texttt{OLD} replaces \texttt{PRold}, but serves a very similar purpose.
Instead of the values of \texttt{PR} in the previous generation, it now
contains the value of \texttt{PR[w]} of the last time the loop body was executed for \texttt{w}.

Note that this specification quite naturally ended up being a ``push-style'' algorithm, pushing changes forward in the graph, rather than pulling them forward.
Another noteworthy change is the addition of the if-statement. Rather than
continuously executing this loop, it is only executed when a change is
produced.
To correct for the missing reinitialization, instead of initializing the values of \texttt{PR} at $\frac{1}{|V|}$, they now need to be initialized at $\frac{1-d}{|V|}$.

Unfortunately, while this specification is much less strict about order, it
still contains a double iteration: first over vertices, and then over edges.
To enable the Forelem framework to fully optimize this specification, we want to impose as little structure as possible.
Fortunately, this can be solved exactly like the first step: with \texttt{OLD}.
By expanding \texttt{OLD} to be per-edge, rather than per-vertex, it is no longer required to do all updates for a vertex at once.
In turn, this enables iterating over edges, getting rid of the imposed push style, and optionally enabling a pull style as well.
The resulting specification is shown in Algorithm~\ref{alg:pagerank}. A
proof of the correctness of this specification as well as its convergence
properties is given in Appendix A.2.

\begin{algpagerank}
\caption{The initial Forelem specification of PageRank.}
\label{alg:pagerank}
\small
\begin{alltt}
\textbf{whilelem} (\tuple{u,v} \(\in\) E) \{
  \textbf{if} (PR[u] != OLD[u,v]) \{
    PR[v] += d*(PR[u]-OLD[u,v])/Dout[u]
    OLD[u,v] = PR[u]
  \}
\}
\end{alltt}
\normalsize
\end{algpagerank}

\section{Transformations and Implementations}
\label{sec:transformations}
The Forelem specifications given in \autoref{alg:kmeans} and
\autoref{alg:pagerank} capture the essence of the applications and are used as
a starting point to derive several implementations, by an automated process.
This automated process functions through the application of sequences of
transformations. In this section, the transformations are described that are
used to derive the final implementations of which the performance is
evaluated in \autoref{sec:evaluation}. Note that more transformations, and
many more implementations are possible. For a detailed description we refer
to \cite{tUPLCPC}, and for the automatic generation of the associated data
structures and the overall automation process we refer to
\cite{tUPLintro,vstrien-2017}.

\subsection{Orthogonalization}
\label{sec:orthogonalization}
The orthogonalization transformation can be used to optimize the order in
which tuples are visited. It introduces an outer loop, which adds an order
to the processing of the tuples. The outer loop selects one or more fields
of the tuples, the inner loop then loops over those tuples in the original
reservoir which contain the selected values for these fields. Thereupon the
reservoir splitting transformation, that will be described next, can be
applied to this outer loop. As a result, tuples are now processed in
particular groups, and reservoir splitting can be performed based on the
values for certain fields of the tuple.

For example, if orthogonalization is applied to k-Means, the outer loop will
iterate over all data
points and the inner loop over all clusters. This can be seen in
\autoref{alg:kmeansortho}. $\texttt{T.x}$ denotes the set of values of field
\texttt{x} that occur in tuples $\texttt{t} \in \texttt{T}$.  Also note that
$\texttt{T.x[y]}$ is notation to select all tuples $\langle
\texttt{m,x}\rangle \in \texttt{T}$ such that $\texttt{y == x}$.

For the case of PageRank, the resulting outer loop iterates over target
vertices, and the inner loop over edges that have said vertex as target.
This results in \autoref{alg:ortho}.

\begin{algkmeans}
\caption{The Forelem specification of k-Means clustering
(\autoref{alg:kmeans}) after using orthogonalization.}
\label{alg:kmeansortho}
\small
\begin{alltt}
\textbf{whilelem} (y \(\in\) T.x)
 \textbf{forelem} (\tuple{m,x} \(\in\) T.x[y])
  \textbf{if} (M[x] != m
      && dist(COORDS[x],M_COORDS[m])
         < dist(COORDS[x],M_COORDS[M[x]])) \{
     M_COORDS[M[x]] =
      (M_COORDS[M[x]]*M_SIZE[M[x]] - COORDS[x])
       / (M_SIZE[M[x]] - 1)
     M_SIZE[M[x]] -= 1
     M_COORDS[m] =
      (M_COORDS[m]*M_SIZE[m] + COORDS[x])
       / (M_SIZE[m] + 1)
     M_SIZE[m] += 1
     M[x] = m
    \}
\end{alltt}
\normalsize
\end{algkmeans}

\begin{algpagerank}
\caption{The Forelem specification of PageRank (\autoref{alg:pagerank})
after applying orthogonalization.}
\label{alg:ortho}
\small
\begin{alltt}
\textbf{whilelem} (w \(\in\) E.v)
  \textbf{forelem} (\tuple{u,v} \(\in\) E.v[w])
    \textbf{if} (PR[u] != OLD[u,v]) \{
        PR[v] = d*(PR[u]-OLD[u,v])*(1/Dout[u])
        OLD[u,v] = PR[u]
    \}
\end{alltt}
\normalsize
\end{algpagerank}

\subsection{Reservoir Splitting}
\label{sec:reservoir-split}
The Forelem framework provides an inherently parallel specification.
Similar to the automatic data partitioning and loop blocking optimization
techniques in optimizing compilers, Forelem can automatically partition the
tuple reservoir to parallelize the execution of forelem and whilelem loops.
Contrary to traditional optimizations like data partitioning and loop
blocking which have to take into account loop iteration data dependencies
and array bounds, within the Forelem framework this partitioning transformation
is very straightforward: as no structure is imposed on the shared spaces,
all iterations are naturally parallel and all tuples have the same
structure.  Given a reservoir $\texttt{R}$ to iterate over, we partition it in
subreservoirs, then iterate over these separately, ideally in parallel. Any
partitioning of $\texttt{R}$ works, as long as
$\bigcup_{\texttt{i}}\texttt{S(R)}_{\texttt{i}} = \texttt{R}$. Note that
because of the use of the union operator, the tuples are allowed to reside
in multiple partitions. Usually a fair partitioning is used, where every
$\texttt{S(R)}_{\texttt{i}}$ has roughly the same size.

If we apply single-value reservoir splitting on the initial Forelem
specification of PageRank (\autoref{alg:pagerank}), this results in
\autoref{alg:lb1}. Note that although this variant is simple, in practice
it will be suboptimal, since it requires synchronization between different
partitions on the various writes to $\texttt{PR}$, see
\autoref{sec:sspace-allocation}. If reservoir splitting is applied to
\autoref{alg:ortho}, this would result in a program specification in which
every $\texttt{PR}$ value has exactly one writer (see
\autoref{alg:o1e1lb1}).

If reservoir splitting is applied to the orthogonalized code of the k-Means
algorithm (\autoref{alg:kmeansortho}), both loops are replaced by:

\small
\begin{alltt}
\textbf{whilelem} (y \(\in\) S(T)\_i.x)
   \textbf{forelem} (\tuple{m,x} \(\in\) S(T)\_i.x[y])
\end{alltt}
\normalsize

\noindent
in case the reservoir splitting is based on single values of the field
\texttt{x}. The reservoir splitting can also be based on a range of values
of field \texttt{x} in this case the two loops are replaced by

\small
\begin{alltt}
\textbf{whilelem} (y \(\in\)
  T.x[ [min(T.x)+i*((max(T.x)-min(T.x))/S),
        min(T.x)
          +(i+1)*((max(T.x)-min(T.x))/S)-1] ])
  \textbf{forelem} (\tuple{m,x} \(\in\) T.x[y])
\end{alltt}
\normalsize

\noindent
in which \texttt{i} ranges over the numbers $\texttt{[0, S-1]}$. The
resulting code of the single-value reservoir splitting on
\autoref{alg:kmeansortho} is shown in \autoref{alg:kmeansorthosplit}. Note
that for this resulting code, each partition only needs the $\texttt{COORDS}$
and $\texttt{M}$ values that apply to its own data points.

\begin{algkmeans}
\caption{The Forelem specification of k-Means clustering
after the application of reservoir splitting on \autoref{alg:kmeansortho}.}
\label{alg:kmeansorthosplit}
\small
\begin{alltt}
\textbf{whilelem} (y \(\in\) S(T)\_i.x)
  \textbf{forelem} (\tuple{m,x} \(\in\) S(T)\_i.x[y])
    \textbf{if} (M[x] != m &&
        dist(COORDS[x],M_COORDS[m])
        < dist(COORDS[x],M_COORDS[M[x]])) \{
      M_COORDS[M[x]] =
        (M_COORDS[M[x]]*M_SIZE[M[x]]
         - COORDS[x]) / (M_SIZE[M[x]] - 1)
      M_SIZE[M[x]] -= 1
      M_COORDS[m] =
        (M_COORDS[m]*M_SIZE[m] + COORDS[x])
         / (M_SIZE[m] + 1)
      M_SIZE[m] += 1
      M[x] = m
    \}
\end{alltt}
\normalsize
\end{algkmeans}

\begin{algpagerank}
\caption{The Forelem specification of PageRank
(Algorithm~\ref{alg:pagerank}) after the application of reservoir
splitting.}
\label{alg:lb1}
\small
\begin{alltt}
\textbf{whilelem} (\tuple{u,v} \(\in\) S(E)\_i)
  \textbf{if} (PR[u] != OLD[u,v]) \{
    PR[v] = d*(PR[u]-OLD[u,v])*(1/Dout[u])
    OLD[u,v] = PR[u]
  \}
\end{alltt}
\normalsize
\end{algpagerank}

\subsection{Localization}
In order to take full advantage of memory hierarchies, the localization
transformation brings the data in shared spaces directly to the tuples in
the reservoir. So, instead of data stored in shared spaces, this data is
included in the fields of the tuples. Where shared space data is initially
stored separate from tuples, localization causes shared space data to be
stored -- or localized -- in the tuples directly.

For k-Means, using the localization transformation (followed by
orthogonalization and reservoir splitting) we obtain the specification shown in
\autoref{alg:kmeanslocal}. A tuple $\langle
\texttt{m,x,c\_x}\rangle$ in $\texttt{T}$ now contains the value of
a data point $\texttt{x}$ and the associated cluster as
$\texttt{c\_x}$. Note that the $\texttt{x}$ now represents the
actual data point, instead of the index at which this data can be found.

For PageRank, the localization transformation will include the
$\texttt{OLD}$ data with the edge data, yielding tuples
$\tupleM{\texttt{u,v,old}}$. See Algorithm~\ref{alg:localisation}.

\begin{algkmeans}
\caption{The Forelem specification of k-Means clustering
after applying localization to \autoref{alg:kmeansorthosplit}.}
\label{alg:kmeanslocal}
\small
\begin{alltt}
\textbf{whilelem} (y \(\in\) S(T)\_i.x)
  \textbf{forelem} (\tuple{m,x,c\_x} \(\in\) S(T)\_i.x[y])
    \textbf{if} (c\_x != m && dist(x,M_COORDS[m])
         < dist(x,M_COORDS[c\_x])) \{
      M_COORDS[c\_x] = (M_COORDS[c\_x]*M_SIZE[c\_x]
            - x) / (M_SIZE[c\_x] - 1)
      M_SIZE[c\_x] -= 1
      M_COORDS[m] = (M_COORDS[m]*M_SIZE[m] + x) 
                     / (M_SIZE[m] + 1)
      M_SIZE[m] += 1
      c\_x = m
    \}
\end{alltt}
\normalsize
\end{algkmeans}

\begin{algpagerank}
\caption{The Forelem specification of PageRank (\autoref{alg:pagerank})
after applying the localization transformation.}
\label{alg:localisation}
\small
\begin{alltt}
\textbf{whilelem} (\tuple{u,v,old} \(\in\) E)
  \textbf{if} (PR[u] != old) \{
    PR[v] = d*(PR[u]-old)*(1/Dout[u])
    old = PR[u]
  \}
\end{alltt}
\normalsize
\end{algpagerank}

\subsection{Tuple Reservoir Reduction}\label{sec:trr}
The tuple reservoir reduction transformation reduces the iterated tuple
reservoir's size by identifying common subsets ($\texttt{C}$) in the
reservoir which can be compacted when initializing the reservoir and
expanded on demand.  In order to guarantee an efficient implementation,
these subsets $\texttt{C}$ of \texttt{T} are only identified if the tuples
corresponding to these subsets $\texttt{C}$ can be enumerated in linear
(constant) time by a simple enumeration function $\texttt{G}_{\texttt{C}}$.

Having identified subsets and their enumeration functions then the tuple reservoir can be reduced by deleting all tuples of a subset and replacing them by a simple stub to the corresponding enumeration function.
Then at execution time this subset is being generated one at a time and the loop body is replicated for each of the tuples corresponding to this subset.

For k-Means, this transformation is not directly applicable. In the case of
PageRank, these subsets $\texttt{C}\sb{\texttt{u}}$ originate from these vertices
$\texttt{u}$ which originally had an outdegree of 0 and consist of all added
tuples \tuple{u,v} (where $\texttt{u} \neq \texttt{v}$). Assuming the
vertices are numbered $\texttt{1}$ to $\texttt{|V|}$, then each subset
$\texttt{C}\sb{\texttt{u}}$ consists of
$\left\{\langle\texttt{u,i}\rangle  \middle|  \texttt{1} \leq \texttt{i} \leq \texttt{|V|}\right\}$
and the enumeration function ends up
being a simple for-loop from $\texttt{1}$ to $\texttt{|V|}$. When this
transformation is applied to \autoref{alg:pagerank} this results in
\autoref{alg:tsr1}.

\begin{algpagerank}
\caption{The Forelem specification of PageRank (\autoref{alg:pagerank}) after
after applying Tuple Reservoir Reduction}
\label{alg:tsr1}
\small
\begin{alltt}
\textbf{whilelem} (\tuple{u,v} \(\in\) E)
  \textbf{if} (PR[u] != OLD[u,v]) \{
    \textbf{if} (v == \$C)
      \textbf{forelem} (w \(\in\) V\textbackslash\{u\})
        PR[w] = d*(PR[u]-OLD[u,v])*(1/Dout[u])
    \textbf{else}
      PR[v] = d*(PR[u]-e.old)*(1/Dout[u])
    OLD[u,v] = PR[u]
  \}
\end{alltt}
\normalsize
\end{algpagerank}

Note that in contrast to the initial specification of the tuple reservoir
$\texttt{E}$, for which for every vertex with outdegree 0 additional tuples
$\langle\texttt{u,v}\rangle$ were created for every $\texttt{v} \neq
\texttt{u}$, by using tuple reservoir reduction all these tuples were
identified as reducible subsets and therefore deleted.  In fact, the initial
expansion of the tuple reservoir was needed to obtain a clean and simple
representation in the Forelem framework --- thereby allowing a cleaner
convergence proof and facilitating other transformations to be applied to
this specification, see below.  Instead of the generation of the Forelem
construct enumerating all the elements of the subset also an arbitrary
element of this subset could have been chosen.  In this case it is important
that the enumeration function can produce an arbitrary element of the subset
in constant time.

\subsection{Shared Space Allocation}
\label{sec:sspace-allocation}
Because the operations in the forelem and whilelem loop structures are
atomic and inherently parallel, the parallel execution of these loop
structures is straightforward
even in the presence of the fact that different tuples can still access
the same address in shared space \texttt{A}. So all these accesses
through the address function $\texttt{F}_{\texttt{A}}$ can be executed
independently and in any order, whether they are writes or reads.
So, when parallelizing these loop structures
the only choice remains whether to distribute or not the
shared spaces involved in these loop structures.  Whenever the parallel
execution is scheduled on a global shared-memory architecture, i.e.
multi-core architecture, the shared space can just reside in this global
memory. If the parallel execution is scheduled on a distributed parallel
computer, i.e. MPI-based HPC cluster architectures, then a decision has to
be made whether the shared space is replicated or distributed among the nodes of
such a cluster. In both cases, any updates done by tuples on shared
space locations in one node have to be communicated at some time to copies of this
location residing on other nodes. For a general allocation scheme of the
shared space In general this can be very inefficient,
therefore the Forelem framework relies on allocation of the shared space, whereby the
allocation is guided by an orthogonalization transformation
followed by the reservoir splitting transformation (see
\autoref{sec:reservoir-split}). This is done in such a way that the
reservoir splitting transformation is determined either directly by the outer
loop or by a ``loop blocking'' transformation of the outer loop that was
introduced by orthogonalization (see \autoref{sec:orthogonalization}). As a
result, we have the following parallel loop structure on a tuple reservoir
\texttt{T} with orthogonalization performed on a field \texttt{x}:
\small
\begin{alltt}
\textbf{distrfor} (p \(\in\) [0, P-1])
  \textbf{whilelem} (y \(\in\)
    T.x[ [min(T.x)+p*((max(T.x)-min(T.x))/P),
          min(T.x)+
           (p+1)*((max(T.x)-min(T.x))/P)-1] ])
\end{alltt}
\normalsize
\noindent
The shared space allocation in this case is being done in such a way that
all shared space addresses being referenced by the tuples of
the whilelem loop are being allocated in the same node. Note that, because
$\texttt{F}_{\texttt{A}}$ is an affine mapping and the shared space
allocation is determined by the values of a single field \texttt{x},
determining which shared addresses are ``shared'' between different nodes
can be computed efficiently by determining the subset of these spaces.

The only issue which remains is how the actual communication is performed.
Note that because of the inherent parallel nature of the whilelem loop,
the updates do not have to be executed instantaneously but can be done at
any time during which the whilelem loop executes.
So, the shared space location at different nodes do not necessarily have to
be consistent and containing the same values at all times. So it can
be the case that there are multiple (even more than two) values residing in
the same shared space location in different nodes. Also, it does not
matter whether one copy is used to update another copy or whether the other
copy updates the first copy.

In order to make the data exchange more efficient the Forelem framework will
try to accumulate multiple updates. This is done in several ways. The first
way (\emph{buffered data exchange}) buffers all updates on each node in such
a way that multiple iterations of the whilelem loop structure are first
executed before initiating this data exchange. The second way (\emph{master
data exchange}) is established by sending all of these buffered updates to
one master node which reduces all the individual updates to one single
update over all copies which is communicated to all participating nodes.
This latter accumulation step can be optimized further in case we have
update statements (of the from \texttt{a = a + 3} or \texttt{b = b / 2}), in
which case multiple updates of the same variable and be first combined
instead of being performed repeatedly. A third way (\emph{indirect data
exchange}) of optimizing this data exchange can be achieved by relating the
update statements with non-shared space data by program assertions in the
original Forelem specification. As an illustration, if we look at k-Means
then the updates on \texttt{M\_COORDS} and \texttt{M\_SIZE} are directly related to
the means reassignment statement, see \autoref{alg:kmeans}. So, if by
program assertion the coupling is made explicit by asserting that
\texttt{assert(M\_SIZE[i] = SUM(i == M[x] for x in T.x))} then the accumulated update
can be replaced by recomputing this information explicitly for any shared
space location of \texttt{M\_SIZE}.

\subsection{Materialization}
\label{sec:materialization}
So far we have been iterating over tuple reservoirs, without specifying the
relevant data structure. Materialization is the first step in the process of
deriving different data structures and associated shared space data
exchanges. In the Materialization step, an initial choice will
be made on how tuples will be retrieved from the tuple reservoir using an
indexing structure. This index structure will identify a unique integer to
every tuple in the reservoir, without fixating the order in which these
tuples receive this index. So, the original iteration of an unordered
reservoir ($\texttt{t} \in \texttt{T}$) will be replaced by a sequential
iteration $\texttt{i} \in \texttt{[0, N-1]}$. As a consequence, every tuple referral in
the loop body is replaced with $\texttt{PT[i]}$. Note that this is just a syntax
change of the loop structure and as such has no consequences for the
execution order of the original forelem or whilelem loop. As an illustration
when materialization is applied to \autoref{alg:ortho}, we get the loop
structure of \autoref{alg:PR-ortho-matr}.

\begin{algpagerank}
\caption{Materialization applied to the initial Forelem Specification of
PageRank (Algorithm~\ref{alg:pagerank}.)}
\label{alg:PR-ortho-matr}
\small
\begin{alltt}
\textbf{whilelem} (i \(\in\) [0,|PE|-1])
  \textbf{if} (PR[PE[i].u] != OLD[PE[i]]) \{
    PR[PE[i].v] = d*(PR[PE[i].u]-OLD[PE[i]])
        *(1/Dout[PE[i].u])
    OLD[PE[i]] = PR[PE[i].u]
  \}
\end{alltt}
\normalsize
\end{algpagerank}

The next step towards actual data structure generation is a concretization
step in which the order of the iterations is fixated and an initial choice
of actual data structures is made. This data structure specification does
not specifically specify whether to use pointer-structured lists, arrays,
dictionaries, etc., but rather specifies the logical structure of the data
structures. For instance, array of structures, arrays of arrays or structures
of arrays. After the concretization phase and at the code generation phase
these logical structures will be replaced by (target code) dependent data
structure declarations, for instance plain arrays, arrays represented as
linked lists, pointer-linked grid structures, STL containers, balanced
trees, etc.

One might wonder why the initial materialization step is not combined with
the concretization step. This choice was made so that the initial
materialization step can be combined with the code transformations as
described before, thereby creating opportunities for deriving unexpected
data structure choices. For instance consider the Forelem specification of
sparse matrix multiplication (see Section 3). Then, if we apply localization of
shared space $\texttt{A}$ on the specification, followed by
orthogonalization on \texttt{i} and \texttt{j} successively,
materialization of tuple reservoir and storing the materialized tuple fields
in separate arrays, we obtain the loop structure:
\small
\begin{alltt}
\textbf{forelem} (j \(\in\) [0,N-1])
  \textbf{forelem} (i \(\in\) [0,N-1])
    \textbf{forelem} (kk \(\in\) [0,|PA[i]|-1]])
      C[i,j] += PA[i][kk]*B[PB[i][kk],j]
\end{alltt}
\normalsize
\noindent
Then after a loop interchange transformation of the two inner loops the
resulting loop structure is:
\small
\begin{alltt}
\textbf{forelem} (j \(\in\) [0,N-1])
  \textbf{forelem} (kk \(\in\) [0,max_i(|PA[i]|])-1])
    \textbf{forelem} (i \(\in\) [0,N-1])
      C[i,j] += PA[i][kk]*B[PB[i][kk],j]
\end{alltt}
\normalsize
\noindent
Now, if concretization is applied to this loop structure, then we obtain a
jagged diagonal/ITPACK data structure implementation for sparse matrix
multiplication~\cite{kincaid-1983}. This resulting data structure would not
have happened if we would have combined the concretization with the
materialization step, because after concretization loop interchange, and
possible other transformations, will be obscured by the fixed logical
structure of the data structures.  This is the prime reason why optimizing
compiler techniques thus far have not been able to transform any matrix
multiplication into this jagged diagonal/ITPACK form. This implementation
relying on jagged diagonal/ITPACK has been very successful in exploiting
long pipelined executions (vector processing) architectures but always have
been developed by hand~\cite{kincaid-1983,bai-2000}.

\subsection{Putting It All Together}
The transformations as described in the previous section can be composed so
that their effect is multiplied. Note that, except for concretization, the
transformations have a Forelem specification as input and produce Forelem
specifications as output, so they are inherently composable. The composition
of multiple transformations allowing different orders of application ---
including re-use of transformations --- leads to many different
implementation of the initial Forelem specification. These implementations
can differ in many aspects: iteration order of the tuples, reservoir
partitioning, generated data structure and the shared space data exchange
that is used.

\subsubsection{k-Means}
As a running example in the previous section, we have already seen the
composition of transformations applied to the initial k-Means specification
that was given in \autoref{alg:kmeans}. Orthogonalization (resulting in
\autoref{alg:kmeansortho}) caused the tuple reservoir to be iterated
point-by-point by the outer loop, introducing an inner loop to evaluate the
distance to each mean for this particular point. This was followed by
reservoir splitting (resulting in \autoref{alg:kmeansorthosplit}), which
partitioned the tuple reservoir enabling parallel execution whereby each
process is executing the loop body for a subset of the tuple reservoir. The
subsequent application of the localization transformation results in a code
(\autoref{alg:kmeanslocal}) in which data from the \texttt{COORDS} and
\texttt{M} shared spaces is included in, or grouped with, the tuples.

We now demonstrate that the application of materialization at different
steps in the transformation chain leads to different data structures to be
generated as well as different data exchange schemes. First
consider \autoref{alg:kmeansorthosplit}. Then, after applying
orthogonalization and materialization we obtain
\autoref{alg:kmeansorthomat}, in which all shared spaces became separate
array structures.

\begin{algkmeans}
\caption{The Forelem specification of k-Means clustering after using
orthogonalization and reservoir splitting (\autoref{alg:kmeansorthosplit})
followed by materialization.}
\label{alg:kmeansorthomat}
\small
\begin{alltt}
\textbf{whilelem} (i \(\in\) [0,|PM|-1])
  \textbf{forelem} (m \(\in\) [0,k-1])
    \textbf{if} (PM[i] != m &&
        dist(PCOORDS[i],PM\_COORDS[m])
        < dist(PCOORDS[i],PM\_COORDS[PM[i]])) \{
      PM\_COORDS[M[i]] =
        (PM\_COORDS[M[i]]*PM\_SIZE[M[i]]
         - PCOORDS[i]) / (PM\_SIZE[M[i]] - 1)
      PM\_SIZE[M[i]] -= 1
      PM\_COORDS[m] =
        (PM\_COORDS[m]*PM\_SIZE[m] + PCOORDS[i])
         / (PM\_SIZE[m] + 1)
      PM\_SIZE[m] += 1
      PM[i] = m
    \}
\end{alltt}
\normalsize
\end{algkmeans}

\noindent
When allocating the shared spaces among distributed memory nodes, the most
natural choice for \autoref{alg:kmeansorthomat} can be either distributed
via buffered and/or indirect data exchange. The straightforward implementation
of buffered data exchange is to perform this data exchange after one or
multiple sweeps of the inner forelem loop.  With respect to the indirect
data exchange the updates on \texttt{PM\_SIZE} can be computed indirectly from
the updates on the \texttt{PM\_COORDS} and \texttt{PM}. See also
\autoref{sec:sspace-allocation}.

As a second materialization possibility, we consider the application of
materialization to \autoref{alg:kmeanslocal}, which is a localized version
of \autoref{alg:kmeans}. Also in this case, the outer loop is replaced with a
loop over an integer interval, but contrary to \autoref{alg:kmeans} the
values in the tuples are not residing in shared memory anymore but are part
of the tuples themselves. So, the tuples in the reservoir also need to be
materialized. Note that for the materialization of the tuples like the
materialization of the shared space an explicit data structure is not chosen at this
step, but rather an indexing structure is set up that is
indexed with integers. As a result we obtain \autoref{alg:kmeanslocalmat}.
Observe that \texttt{DATA} and \texttt{M} are not stored as arrays in this
case, due to the application of localization. In the concretization step,
this materialized version of the tuple reservoir will be transformed to an
array of structures in which each structure contains a
vector to store \texttt{x} and integer to store \texttt{c\_x}.
The allocation of the shared spaces for this variant
are similar to \autoref{alg:kmeansorthomat} and therefore not further
discussed.

\begin{algkmeans}
\caption{The Forelem specification of k-Means clustering after using
orthogonalization, reservoir splitting and localization
(\autoref{alg:kmeanslocal}) followed by materialization.}
\label{alg:kmeanslocalmat}
\small
\begin{alltt}
\textbf{whilelem} (i \(\in\) [0,|PT|-1])
  \textbf{forelem} (m \(\in\) [0, k-1])
    \textbf{if} (PT[i].c\_x != m &&
        dist(PT[i].x,PM\_COORDS[m])
        < dist(PT[i].x,PM\_COORDS[PT[i].c\_x])) \{
      PM\_COORDS[PT[i].c\_x] =
       (PM\_COORDS[PT[i].c\_x]*PM\_SIZE[PT[i].c\_x]
        - PT[i].x) / (PM\_SIZE[PT[i].c\_x] - 1)
      PM\_SIZE[PT[i].c\_x] -= 1
      PM\_COORDS[m] = (PM\_COORDS[m]*PM\_SIZE[m]
        + PT[i].x) / (PM\_SIZE[m] + 1)
      PM\_SIZE[m] += 1
      PT[i].c\_x = m
    \}
\end{alltt}
\normalsize
\end{algkmeans}

\subsubsection{PageRank}
Up till now, we have only seen the application of individual transformations
to the initial PageRank specification. In this section, we will consider
compositions of these transformations. For k-Means, we first applied
orthogonalization, followed by reservoir splitting. If this is done for the
initial PageRank specification (\autoref{alg:pagerank}) this results in
\autoref{alg:o1e1lb1}.
Since $\texttt{OLD}$ is always indexed using $\texttt{e} \in \texttt{E}$,
$\texttt{OLD}$ can be made part of the tuple to avoid repeated separate shared
spaces access. This can be achieved with the localization transformation,
resulting in \autoref{alg:o1e1l1lb1}.

\begin{algpagerank}
\caption{The Forelem specification of PageRank (\autoref{alg:pagerank})
after the application of orthogonalization and reservoir splitting.}
\label{alg:o1e1lb1}
\small
\begin{alltt}
\textbf{forelem} (w \(\in\) S(V)\_i) \{
  \textbf{forelem} (\tuple{u,v} \(\in\) E.v[w]) \{
    \textbf{if} (PR[u] != OLD[u,v]) \{
        PR[v] = d*(PR[u]-OLD[u,v])*(1/Dout[u])
        OLD[u,v] = PR[u]
    \}
  \}
\}
\end{alltt}
\normalsize
\end{algpagerank}

\begin{algpagerank}
\caption{The Forelem specification of PageRank (\autoref{alg:pagerank}) after
the application of orthogonalization, localization and reservoir splitting,
in order.}
\label{alg:o1e1l1lb1}
\small
\begin{alltt}
\textbf{whilelem} (w \(\in\) S(V)_i) \{
  \textbf{forelem} (\tuple{u,v,old} \(\in\) E.v[w]) \{
    \textbf{if} (PR[u] != old) \{
      PR[v] = d*(PR[u]-old)*(1/Dout[u])
      old = PR[u]
    \}
  \}
\}
\end{alltt}
\normalsize
\end{algpagerank}

This implementation can now be materialized, such that an index structure is
introduced to access the tuple data.  The result is shown in
\autoref{alg:o1m1m1ts1l1n1n1l1lb1}.

\begin{algpagerank}
\caption{The Forelem specification of PageRank in
Algorithm~\ref{alg:o1e1l1lb1}, after application of materialization.}
\label{alg:o1m1m1ts1l1n1n1l1lb1}
\small
\begin{alltt}
\textbf{whilelem} (v \(\in\) [0,|V|-1]) \{
  \textbf{forelem} (i \(\in\) [0,|PE[v]|-1]) \{
    \textbf{if} (PR[PE[v][i].u] != PE[v][i].old) \{
      PR[v] = d*(PR[PE[v][i].u]-PE[v][i].old)
          *(1/Dout[PE[v][i].u])
      PE[v][i].old = PR[PE[v][i].u]
    \}
  \}
\}
\end{alltt}
\normalsize
\end{algpagerank}

The shared space data exchange scheme that is generated depends on the
distribution of the tuples and shared spaces that is implied by the
transformations that have been performed on a specification. For the case of
\autoref{alg:o1m1m1ts1l1n1n1l1lb1} only shared space \texttt{PR} is written,
all other shared spaces are only read. From analysis follows that
\texttt{PR} is only written by a subscript \texttt{v}, which has only been
distributed to a single process during reservoir splitting, so all writes
are local. Because other processes will read this value, \texttt{PR} must be
kept current. The buffered data exchange scheme that is generated for
PageRank is the same as for k-Means: each process buffers the changed values
and communicates these during a data exchanged. Initially, this data
exchange will take place after each iteration of the inner loop, but again
for efficiency reasons these changed values can be collated such that the
data exchange is only performed after one (or multiple) sweeps of the inner
forelem loop.

Finally, note that in addition to the compositions of transformations that
have been described above, also tuple reservoir reduction as described in
Section~\ref{sec:trr} can be applied to all resulting algorithms.

\section{Experiments}
\label{sec:evaluation}
To evaluate the performance of the derived k-Means clustering and PageRank implementations we ran several experiments, using two implementations from the BigDataBench benchmark as a baseline for each algorithm. We will first briefly discuss these implementations in \autoref{sec:kmeansbigdatabench} and \autoref{sec:pagerankbigdatabench}, for k-Means clustering and PageRank respectively. The experimental setup is explained in \autoref{sec:expsetup} and the results are given in \autoref{sec:results}.

\subsection{k-Means Baseline Implementations}
\label{sec:kmeansbigdatabench}
For obtaining state-of-the-art codes we use as a repository the BigDataBench
benchmark \cite{Wang14}, which contains several parallel implementations of the k-Means clustering algorithm. For our first baseline we choose to use the Hadoop implementation, which uses the implementation included in the Apache Mahout project \cite{mahout}. Hadoop provides a way to easily parallelize existing algorithms, which is something the Forelem framework also wishes to achieve. We will refer to this implementation as the Hadoop\_Mahout implementation. 

Since the final implementations generated using the Forelem framework will
use C/C++ code and MPI, we also use the C/C++ MPI implementation from the
BigDataBench benchmark, which originated from Northwestern University and
was written by W.-K. Liao \cite{mpiNW}. It takes a more traditional approach
to parallelizing k-Means clustering, first all processes reassign the data
points in parallel, then they recalculate the cluster centers in parallel.
We will refer to the C/C++ MPI implementation as Kmeans\_MPI. 

\subsection{PageRank Baseline Implementations}
\label{sec:pagerankbigdatabench}
The PageRank Hadoop implementation in the BigDataBench
benchmark~\cite{Wang14} is the one as implemented in the
Pegasus~\cite{Kang09} project.  It is a fairly straight-forward MapReduce
version of the original algorithm.  Each iteration evaluates the
contributions of each edge, then all those contributions are summed up to
produce the next PageRank values.  The implementation stops when the change
between iterations in smaller than a given error bound $\epsilon$.
In~\cite{Page99} empirical evidence is given this algorithm tends to
converge in a small number of iterations.  Throughout this section, we will
refer to this implementation as the PageRank\_Hadoop implementation

Like for k-Means, BigDataBench also comes with a C++/MPI implementation of
PageRank, which is taken from Indiana University's Parallel BGL~\cite{BGL}.
We will refer to this implementation as the PageRank\_MPI implementation.

\subsection{Experimental Setup}
\label{sec:expsetup}
For the k-Means experiments we have chosen to write a random data generator to
exclude any bias towards initial distributions or other artifacts. The
generator is given the total number of data points to generate, the
dimension of the desired data points and the number of clusters to generate
the data in. It first generates the intended cluster centers using a uniform
distribution in the interval $[0,10]$ and a standard deviation for
each cluster, uniform random in the interval
$[\tfrac{10}{16},\tfrac{10}{8}]$. To generate a data point, the generator
first uniform randomly chooses a cluster to assign it to, then uses a normal
distribution with the generated center as a mean and the generated standard
deviation. Note that it is possible for coordinates of the generated data
points to fall outside the interval $[0,10]$. All data sets used in the
experiment contained data points of dimension 4, generated in 4 clusters.
For each implementation the data is stored in ASCII format, since the format
for the Kmeans\_MPI and Hadoop\_Mahout implementation differ slightly the
Forelem implementations can read both formats. 

For the experiments with the PageRank implementations we use the data
generator provided by the BigDataBench benchmark~\cite{Wang14}.
This generator uses parameters derived from a Google webgraph to create Kronecker graphs of the requested size.
Due to the generation method a very small amount of vertices may not be connected, but this poses no problems for any of the used implementations.

Note that the Kmeans\_Hadoop implementation uses a convergence delta to determine whether the process has converged. If the change in the cluster centers during an iteration is less than this convergence delta, the calculation terminates. To allow a fair comparison, this convergence delta was added to Implementation 1 to 4. Similarly, the Kmeans\_MPI implementation uses a threshold to determine convergence. If the fraction of data points that are switched to a different cluster center during an iteration is less than the given threshold, calculation terminates. This was also added to the Forelem implementations of k-Means. 

The experiments ran on (up to) 16 nodes of the DAS-4 cluster \cite{das4}. A
node in this cluster consists of 2 CPU sockets, each containing a 4-core CPU
with Hyper-Threading. This yields a total of 8 physical cores and 16 virtual
cores per node. So, up to a total of 256 threads were run in parallel.

Throughout this section, we will refer to the implementation of k-Means
clustering corresponding to \autoref{alg:kmeansortho} using the
buffered data exchange scheme as Kmeans\_1. Implementation Kmeans\_2
also corresponds to \autoref{alg:kmeansortho}, but uses the indirect data
exchange scheme scheme. Similarly, Implementation Kmeans\_3 and Implementation
Kmeans\_4 correspond to \autoref{alg:kmeanslocal} and use the indirect data
exchange scheme and buffered data exchange scheme respectively. Similarly,
for the PageRank implementations, we will refer to the implementation
corresponding to \autoref{alg:lb1} as PageRank\_1, the implementation
corresponding to \autoref{alg:o1m1m1ts1l1n1n1l1lb1} as PageRank\_2,
implementation PageRank\_3 corresponds to \autoref{alg:o1e1l1lb1} and
implementation PageRank\_4 corresponds to \autoref{alg:o1e1lb1}. All these
implementations use buffered data exchange.

\subsection{Results}
\label{sec:results}
\begin{figure}
\scalebox{.7}{\input{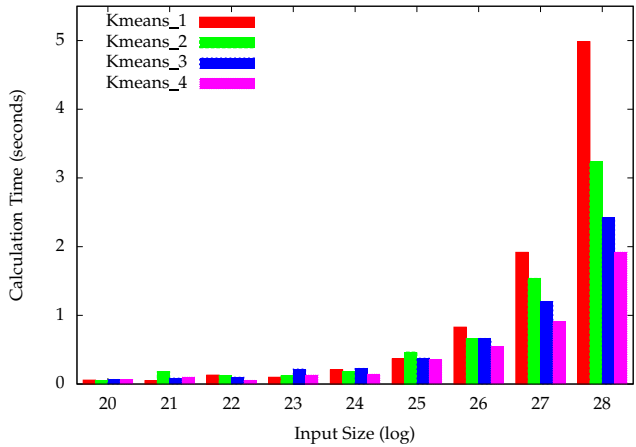}}
\caption{The calculation time the Forelem k-Means implementations, using 64 threads and a convergence delta of $0.0001$.}
\label{fig:bar-calc}
\end{figure}

For the first experiment, we ran the Forelem k-Means implementations using 64 threads for data sets containing $2^{20}$ to $2^{28}$ data points, the results are shown in \autoref{fig:bar-calc}. These results show that applying localization, as is done for the Kmeans\_3 and Kmeans\_4 implementations, decreases the calculation time. Similarly, for larger data sets, the calculation time is decreased by using the derived communication scheme instead of the recalculation communication scheme, as shown by implementations Kmeans\_2 and Kmeans\_4 perfoming better than implementations Kmeans\_1 and Kmeans\_3 respectively. Both effects become more apparent on the larger data sets. Note that the calculation time shown excludes the time needed for input and output. We focus our experiments on the part of the code that was specified and optimized in the Forelem framework: time is measured from the start of the initialization of the cluster centers until the execution of the whilelem loop terminates. 

\begin{figure}
	\scalebox{.7}{\input{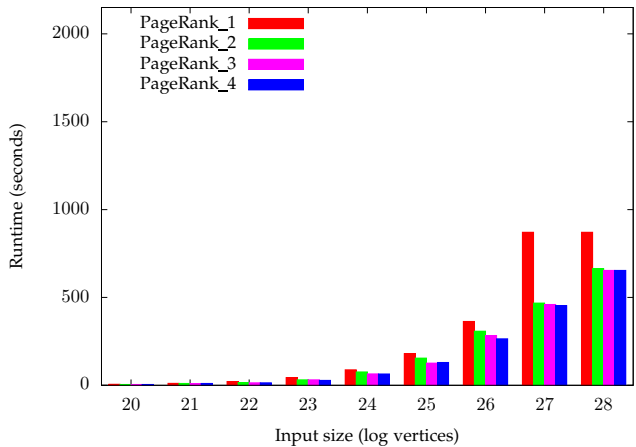}}
	\caption{The runtime of the Forelem PageRank implementations using 64 threads. Hadoop was left out to improve legibility.}
	\label{fig:bestbar}
\end{figure}

\begin{figure}
	\scalebox{.7}{\input{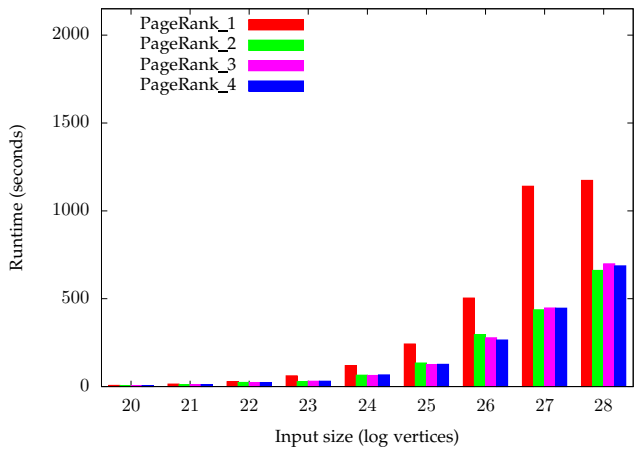}}
	\caption{The runtime of the Forelem PageRank implementations using 128 threads.}
	\label{fig:bestcase}
\end{figure}

Similarly, we ran the Forelem PageRank implementations using 4 threads per node and 8 threads per node (leading to a total of 64 and 128 threads respectively) for data sets containing approximately $2^{20}$ to $2^{28}$ vertices, the results of which are shown in \autoref{fig:bestbar} and \autoref{fig:bestcase}. These results also show that the applied transformations result in an optimization of the final implementations, since the PageRank\_1 implementation is clearly outperformed by the other three implementations. 
In most cases the other three implementation perform roughly the same, with a deviation in the 128-thread case.
Due to its smaller memory footprint, PageRank\_2 scales better than the other implementations when the memory bus gets close to saturation.

\begin{figure}
\scalebox{.7}{\input{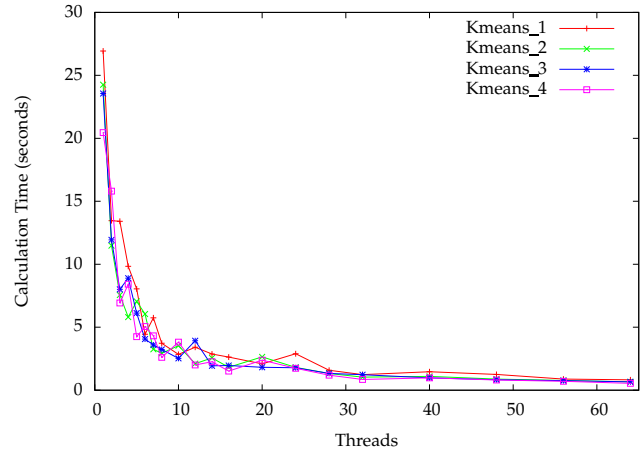}}
\caption{The calculation time of the Forelem k-Means implementations for varying numbers of threads, using a convergence delta of $0.0001$.}
\label{fig:graph-threads}
\end{figure}

\begin{figure}
	\scalebox{.7}{\input{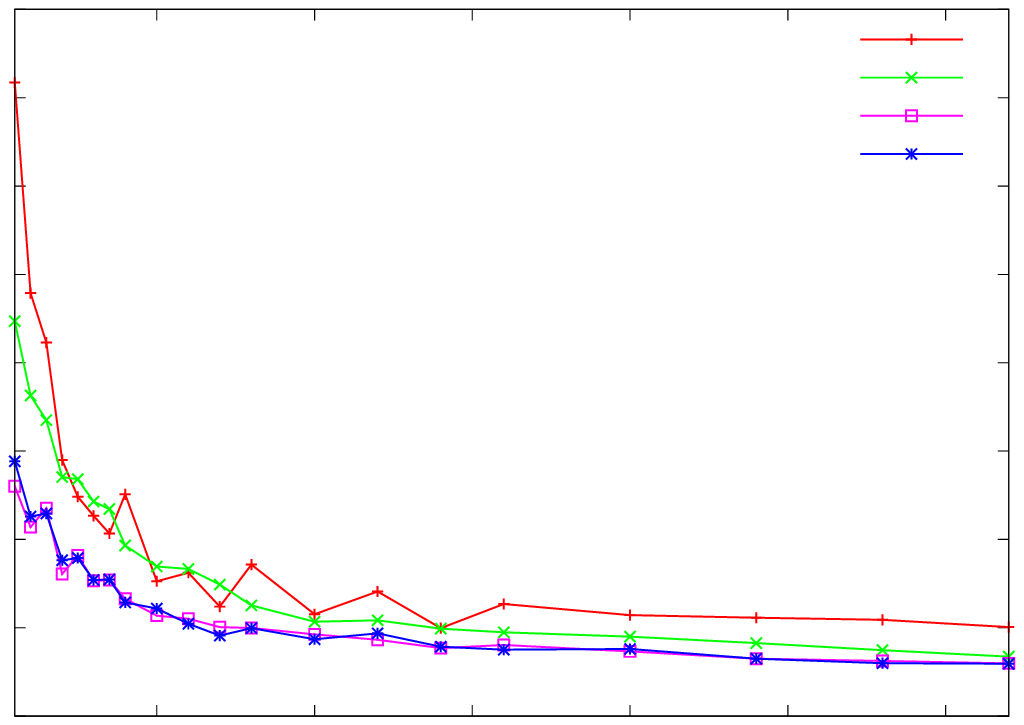}}
	\caption{The runtime of the Forelem PageRank implementations for varying numbers of threads.}
	\label{fig:threadline}
\end{figure}

In a second experiment, the Forelem implementations were run on configurations containing different numbers of threads and the data sets containing $2^{26}$ data points or vertices. The calculation times for the four k-Means implementations are shown in \autoref{fig:graph-threads}, the results for the four PageRank implementations are shown in \autoref{fig:threadline}. Note that when different configurations would yield the same number of threads, the configuration using the lowest number of nodes was used. The results for the k-Means implementations show that when the number of threads double, the calculation time becomes roughly half of the calculation time, thus showing that the implementations scale very well. The PageRank implementations achieve similar results. Due to the range needed to show all results in \autoref{fig:graph-threads} and \autoref{fig:threadline}, the graphs may appear to approach a limit for the higher number of threads, but it in fact they continue to go down at a similar rate as before. From 32 to 64 threads, the calculation time of the k-Means implementations decreases with a factor 1.6 on average. For PageRank, this is a factor 1.3 on average.

\begin{figure}
\scalebox{.7}{\input{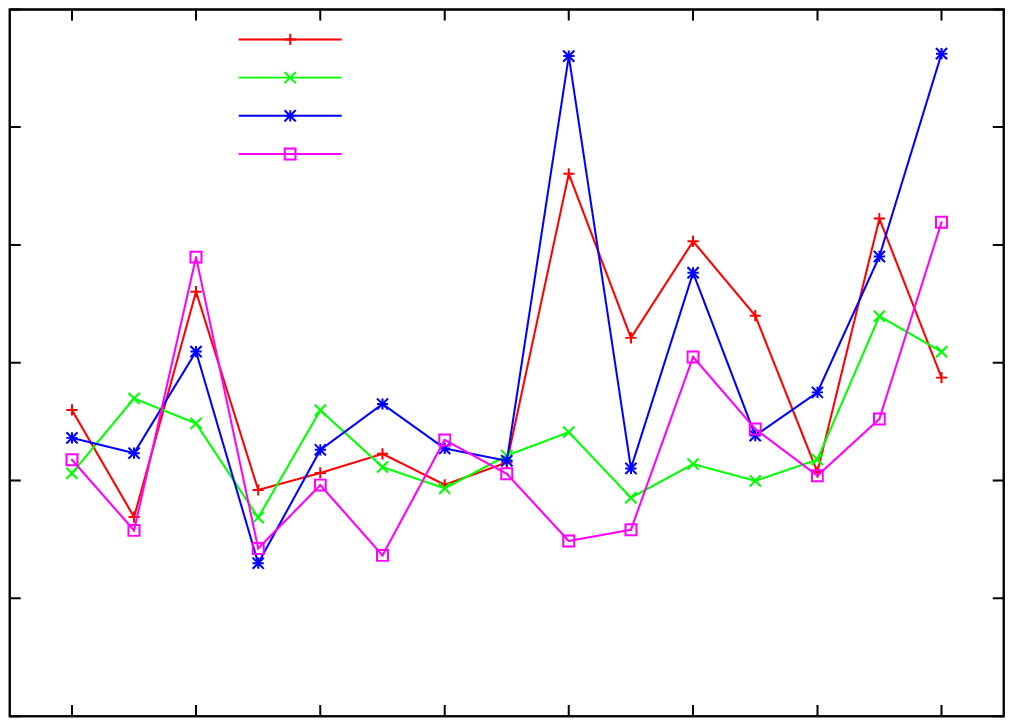}}
\caption{The calculation time of the Forelem k-Means implementations for different dimensions, using 64 threads and a convergence delta of $0.0001$.}
\label{fig:graph-dimension}
\end{figure}

Finally, to further investigate the behaviour of the Forelem k-Means
implementations we also ran an experiment using data sets with different
dimensions and numbers of clusters. Both experiments were run on data sets
of size $2^{26}$. The results when running on data sets with $\texttt{k=4}$ and different dimensions are shown in \autoref{fig:graph-dimension}. These results show that the calculation time slightly increases when the dimension does, which is due to the increase in operations needed to calculate the Euclidean distance and recalculate the cluster centers. However, the increase in calculation time is very small compared to the increase in dimension: an increase of a factor 8 in dimension only results in an increase of about a factor 2 in calculation time. 

\begin{figure}
\scalebox{.7}{\input{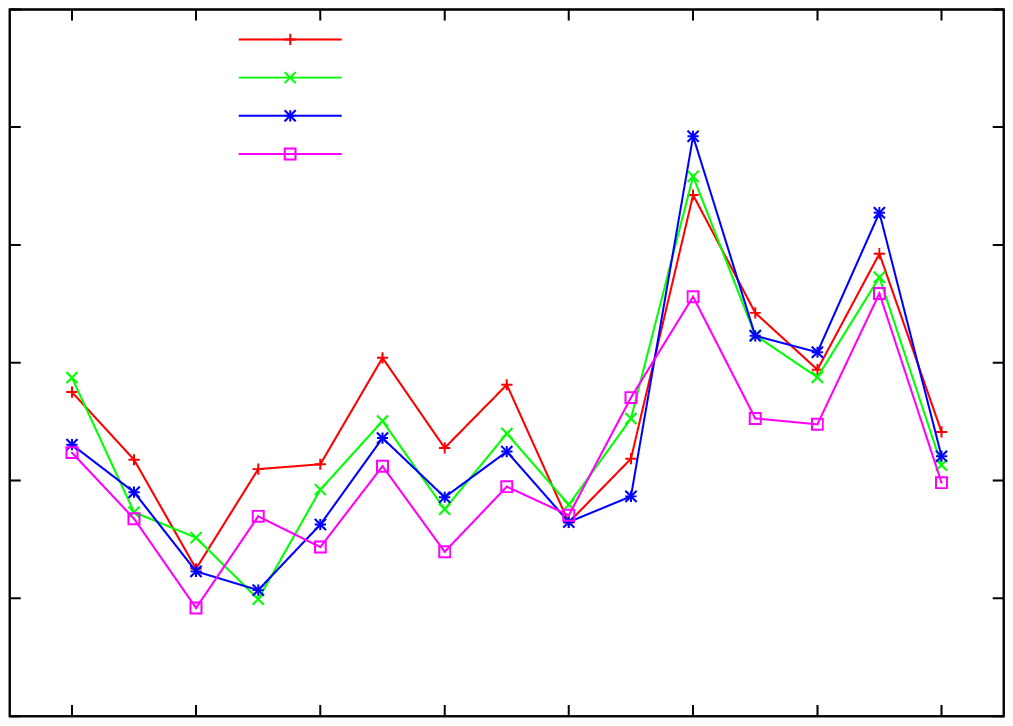}}
\caption{The calculation time of the Forelem k-Means implementations for different $k$, using 64 threads and a convergence delta of $0.0001$.}
\label{fig:graph-clusters}
\end{figure}

The results for data sets with dimension 4 and different numbers of clusters are shown in \autoref{fig:graph-clusters}. Similar to the experiment with different dimensions, the calculation time appears to increase slightly as the number of clusters increases. This is due to an increased amount of information needing to be communicated, when the processes communicate the cluster centers. However, since the frequency of this communication does not increase, only the length of the messages does, the increase in calculation time is small. 

From these first experiments we can conclude that the Forelem framework implementations scale well. The additional optimizations applied in the Kmeans\_4 and PageRank\_2, PageRank\_3 and PageRank\_4 implementations indeed improve the performance, given that Kmeans\_4 is the best performing implementation in almost all cases and PageRank\_1 the worst. To get a better understanding of the performance of the Forelem implementations, we will use the Kmeans\_MPI, Kmeans\_Hadoop, PageRank\_MPI and PageRank\_Hadoop implementations as baselines. 

For the comparison with the Kmeans\_MPI implementation, we note that to allow a fair comparison we will measure only the time taken by the core calculation, similar to how we measure the time for the Forelem k-Means implementations as noted before. The time measurement is taken from the moment the Kmeans\_MPI implementation calls the function that will execute the iterations, until the moment the process has converged. 

\begin{figure}
\scalebox{.7}{\input{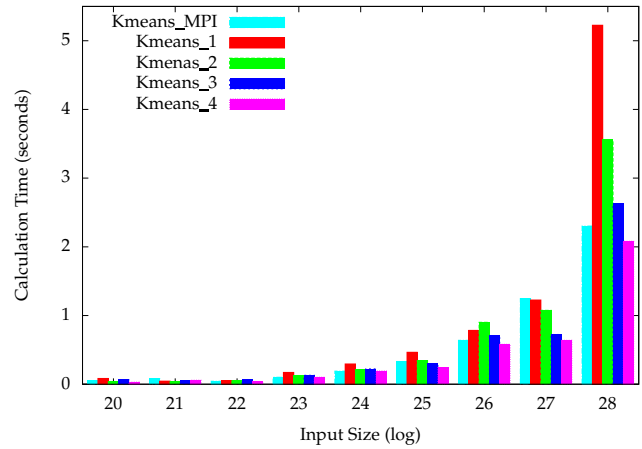}}
\caption{The calculation time of the Forelem k-Means implementations and Kmeans\_MPI implementation, using 64 threads and threshold $0.0001$.}
\label{fig:bar-mpi}
\end{figure}

The performance of the Forelem k-Means implementations and the Kmeans\_MPI implementation on data sets of different sizes are shown in \autoref{fig:bar-mpi}. It should be noted that switching to a threshold of 0.0001 instead of a convergence delta caused outliers. For about 15\% of the runs the number of iterations used became far greater then normally seen (up to 490 iterations in a single run, where runs with 3 to 10 iterations were normal). The Kmeans\_MPI implementation also exhibited this behaviour. These outliers were excluded from the results shown. 

The results show that the performance of Forelem k-Means implementations is close to the performance of the Kmeans\_MPI implementation. The Kmeans\_1 implementation, the slowest of the four Forelem implementations, proved to be slower than the Kmeans\_MPI implementation. The fastest Forelem implementation, Kmeans\_4, proved to be faster for seven out of nine input sizes. For the remaining two input sizes, the Kmeans\_4 implementation was just slightly slower (in the order of 1\%). Note that the comparison is done with a hand-coded MPI version whilst our implementations concern codes which were constructed by an automated process.
 
\begin{figure}
	\scalebox{.7}{\input{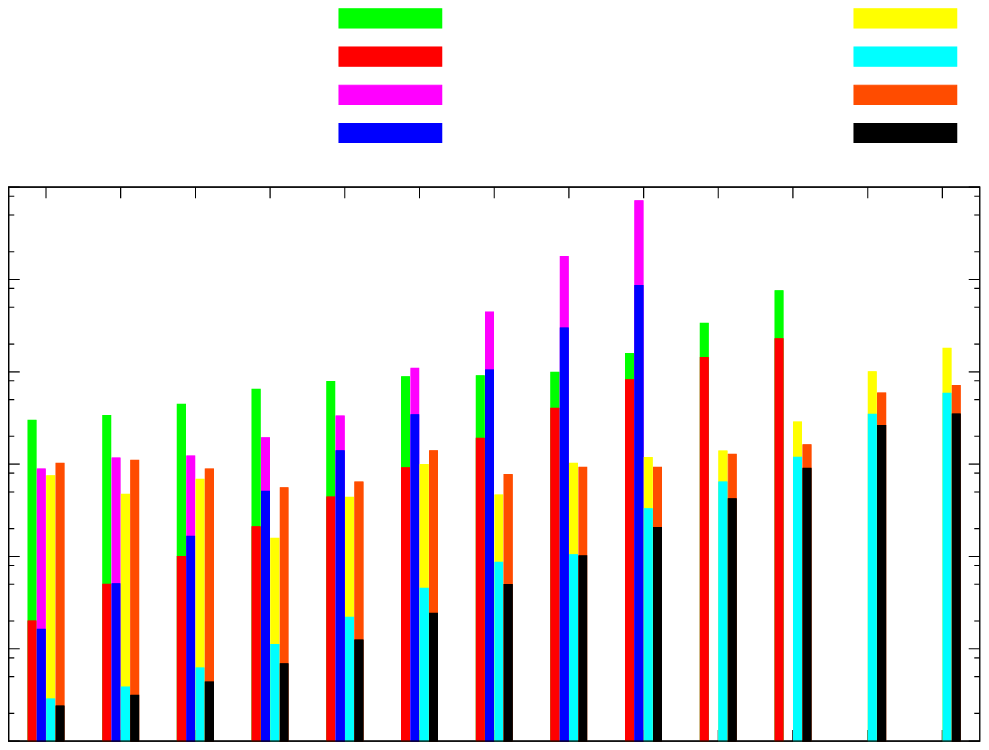}}
	\caption{The worst and best runtimes of PageRank\_MPI, compared to a naive implementation of PageRank in Forelem, and PageRank\_1 and PageRank\_4.}
	\label{fig:mpi-resultaten}
\end{figure}

The performance of the Forelem PageRank implementations was compared to the performance of the PageRank\_MPI implementation, the results of which are shown in \autoref{fig:mpi-resultaten}.
On the smallest graphs, all of our implementations outperform PageRank\_MPI.
In the other cases, PageRank\_MPI only outperforms the most naive of Forelem implementations, and all transformed implementations (PageRank\_1 and PageRank\_4 are shown) significantly outperform PageRank\_MPI.
Of particular note is the memory consumption: on graphs larger than $2^{18}$
vertices, the naive Forelem implementation and PageRank\_MPI consumes too much memory to calculate a result, hence their omission in other experiments.

\begin{figure}
\scalebox{.7}{\input{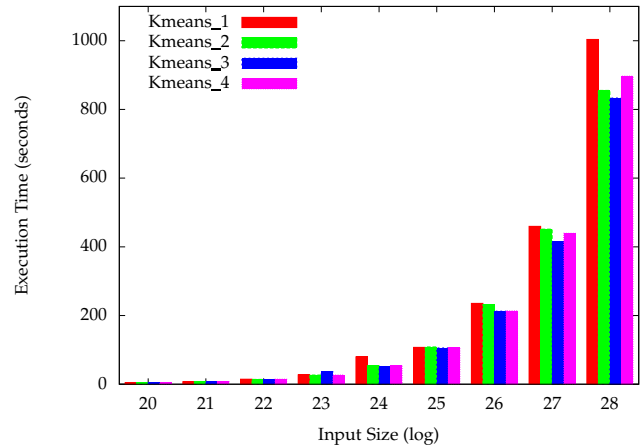}}
\caption{The execution time of the Forelem k-Means implementations, using 64 threads and a convergence delta of $0.0001$.}
\label{fig:bar-exec}
\end{figure}

\begin{figure}
\scalebox{.7}{\input{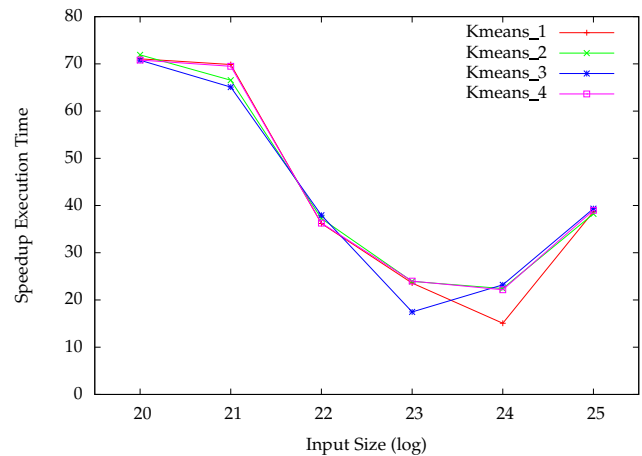}}
\caption{The speedup of the Forelem k-Means implementations compared to the Hadoop\_Mahout implementation, using a convergence delta of $0.0001$.}
\label{fig:graph-hadoop}
\end{figure}

Finally, a comparison with the two Hadoop baselines was made.
\autoref{fig:bar-exec} shows the execution times of the Forelem k-Means
implementations and \autoref{fig:graph-hadoop} shows the speedup of the
Forelem implementations compared to the Kmeans\_Hadoop implementation for
various input sizes. The Kmeans\_Hadoop implementation was given a maximum
number of iterations of 10. Note that the Forelem k-Means implementations
are between 20 to 70 times faster than the Kmeans\_Hadoop implementation.
While the Kmeans\_Hadoop implementation first becomes more efficient
compared to the Forelem implementations as the data size increases, it
becomes less efficient for larger data sizes. However, since for most
implementations this effect is only shown for the largest data set, this may
be coincidental (due to the randomness of the initialization and its
influence on performance). It was not possible to run the Kmeans\_Hadoop
implementation for a data set larger than $2^{25}$ data points, because it
ran out of memory. 

\begin{figure}
	\scalebox{.7}{\input{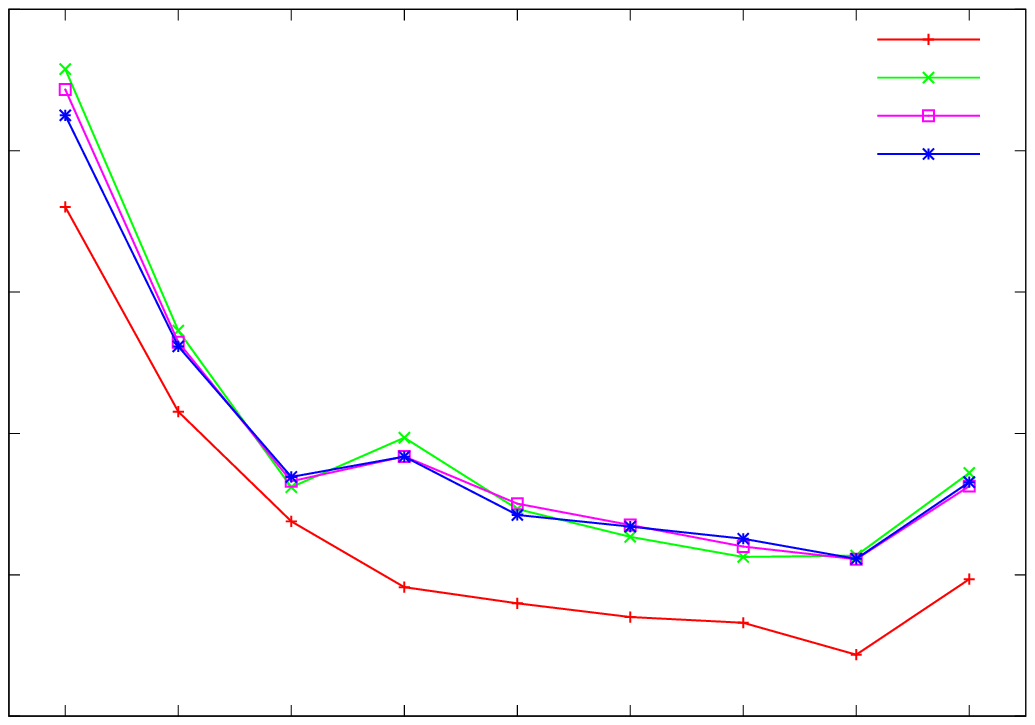}}
	\caption{Speedup of the Forelem PageRank implementations compared to Hadoop using 128 threads.}
	\label{fig:worstline}
\end{figure}

Similarly, we compare the performance of the Forelem PageRank implementations to the performance of the PageRank\_Hadoop implementation, the results of which are shown in \autoref{fig:worstline}.
The figure clearly shows that all Forelem-based implementations outperform the original benchmark implementation.
As expected from the previous results, the performance of Implementations 2, 3 and 4 are closely tied, whereas PageRank\_1 performs the worst.
The minimum speedup achieved by Implementations 2, 3 and 4, at $2^{27}$ vertices, is approximately a factor 60.
Interestingly, for even larger datasets the speedup of these three implementations starts to increase again.
It is likely that this is caused by the fact that for larger datasets, the I/O performed by Hadoop to write intermediate results to disk is becoming a larger and larger bottleneck.

The decrease in speedup that is seen from an input size of $2^{20}$ to $2^{23}$ for all implementations demands an explanation.
We have observed that for smaller datasets Hadoop is not able to fully exploit the available parallelism in the system.
For example, even though 16 nodes are available, Hadoop only performs the mapping task on 7--9 nodes.
The Forelem-based implementations always use all 16 nodes, even for smaller datasets, resulting in significantly faster runtimes.
As the size of the input grows, Hadoop is able to make more efficient use of all available 16 nodes, catching up slightly.

\section{Conclusions}
\label{sec:conclusions}
The Forelem framework is based on inherently parallel specifications. This
demands that a specification process reduces an existing algorithm to its
core idea. By doing so, all computational steps are being desynchronized,
allowing for an automatic translation process which takes into account all
possible computation orders and communication structures, leading to highly
efficient parallel implementations. In this paper we have seen that
through the Forelem framework the application of a sequence of
transformations to a simple specification of an algorithm leads to the
automated derivation of highly efficient implementations. The performance of
these implementations is shown to be superior to the Hadoop baseline
implementations, being approximately 40 to 60 times faster. Also, these
implementations are more efficient than state-of-the-art, handwritten
MPI/C++ implementations. Future work will include the further automation of
this process, and the demonstration of the effectiveness of this framework
on various other examples and algorithms.

\IEEEtriggeratref{4}
\bibliographystyle{IEEEtran}



\end{document}